\documentclass[journal,doublecolumn]{IEEEtranTCOM}
\usepackage{array,cite}
\usepackage{graphicx,epsfig,subfigure}
\usepackage{amsthm}
\usepackage{amsmath}
\usepackage{mdwmath}
\usepackage{array}
\usepackage{subeqnarray}
\usepackage{stfloats}
\usepackage{algorithm,algorithmic}
\usepackage{xcolor}
\usepackage{amssymb}

\begin{document}
\title{Learn to Sense: a Meta-learning Based Sensing and Fusion Framework for Wireless Sensor Networks}
\author{
Hui~Wu, Zhaoyang~Zhang$^{\dagger}$, Chunxu~Jiao, Chunguang~Li, and Tony~Q.S.~Quek
\thanks{Copyright (c) 20xx IEEE. Personal use of this material is permitted. However, permission to use this material for any other purposes must be obtained from the IEEE by sending a request to pubs-permissions@ieee.org. A preliminary version of this work was presented at the 10th International Conference on Wireless Communications and Signal Processing (WCSP 2018) and was published in its Proceedings (DOI: 10.1109/WCSP.2018.8555926). Corresponding author: Zhaoyang Zhang (Email: ning\_ming@zju.edu.cn).}
\thanks{This work was supported in part by National Natural Science Foundation of China under Grant 61725104 and 61631003, and Huawei Technologies Co., Ltd, under Grant YBN2018115223.}
\thanks{H.~Wu and C.~Jiao were with College of Information Science and Electronic Engineering, Zhejiang University (ZJU), Hangzhou 310027, China, and are now with Huawei Technologies Co., Ltd, Shanghai, China. Z.~Zhang and C.~Li are with the College of Information Science and Electronic Engineering, Zhejiang University (ZJU), Hangzhou 310027, China. Tony~Quek is with the ISTD Pillar, Singapore University of Technology and Design (SUTD). The authors are also with the ZJU-SUTD IDEA Center of Network Intelligence.}
}

\maketitle

\begin{abstract}
 Wireless sensor networks (WSN) acts as the backbone of Internet of Things (IoT) technology. In WSN, field sensing and fusion are the most commonly seen problems, which involve collecting and processing of a huge volume of spatial samples in an unknown field to reconstruct the field or extract its features. One of the major concerns is how to reduce the communication overhead and data redundancy with prescribed fusion accuracy. In this paper, an integrated communication and computation framework based on meta-learning is proposed to enable adaptive field sensing and reconstruction. It consists of a stochastic-gradient-descent (SGD) based base-learner used for the field model prediction aiming to minimize the average prediction error, and a reinforcement meta-learner aiming to optimize the sensing decision by simultaneously rewarding the error reduction with samples obtained so far and penalizing the corresponding communication cost. An adaptive sensing algorithm based on the above two-layer meta-learning framework is presented. It actively determines the next most informative sensing location, and thus considerably reduces the spatial samples and yields superior performance and robustness compared with conventional schemes. The convergence behavior of the proposed algorithm is also comprehensively analyzed and simulated. The results reveal that the proposed field sensing algorithm significantly improves the convergence rate.

\end{abstract}


\section{Introduction}

 Internet of Things (IoT) is one of the most promising technologies that has arisen for decades. Through connecting massive communication terminals together, it provides ubiquitous access to almost everything in the world. Among the IoT techniques, wireless sensor network (WSN) is regarded as the backbone due to its capability of collecting, storing, querying and understanding raw sensor data. For instance, the automated switching control of street lamps depends on the monitoring of light intensity via light sensors deployed across the region.

 Advanced and intelligent WSN-based IoT has attracted a lot of research interest. Generally, sensing and fusion are two basic problems. In many applications like environment monitoring, a huge volume of spatial samples are to be collected and processed by the fusion center (FC) to extract some field features or reconstruct the field. In such \emph{field sensing and reconstruction} scenarios, to ensure rapid, accurate and efficient fusion, it is often required to deploy either many less-capable sensors or less sensors each of which capable of collecting many spatial samples, especially when the area of interest is relatively large. Whichever case it is, in addition to the hardware cost of the sensors and the computation cost at the FC, another cost, i.e., the intensive communication between all the sensors and the FC, has become one of the major concerns in system design and realization as spectrum and/or energy resources become stringent. In this regard, how to redesign the sensing, communication and computing processes of a WSN, so as to downsize the dataset and reduce the communication cost with prescribed fusion accuracy, have attracted much attention from both academia and industry.


Much research effort has been devoted to the trade-off between energy consumption and data fusion accuracy for specific field sensing and reconstruction problems. The goal is to design energy-efficient algorithms that effectively reduce the communication cost while maintaining a desired sensing quality. One line of such investigation is to conduct the offline sensor selection based on some information-theoretic criteria before gathering the data. For example, \cite{Joshi2009Sensor,Ghassemi2011Separable,Sebastiani2000Maximum,Paninski2005Asymptotic} mainly focus on the optimal $k$-out-of-$n$ sensor selection problem. Among them, \cite{Sebastiani2000Maximum} exploits the marginal entropy of the sample and \cite{Paninski2005Asymptotic} tends to optimize the mutual information between the unknown system state and the stochastic output samples to make the choice. It is seen that the sensors are filtered before gathering new measurements in these algorithms. As a result, the choices have to be made based on some prior knowledge of the sensing target. Also, it is worth noting that the offline algorithms suffer from two serious drawbacks: a) high computational complexity considering the NP-hard combinational selection problem; b) poor adaptation to the sensing target's dynamic changes.


In contrast to the offline approach, another line of approach is to reduce data redundancy in an online fashion through active sampling techniques \cite{Willett2004Backcasting,Nowak2003Boundary,Rui2005Faster,Rui2008Active,Nguyen2015Information,Grasso2016Dynamic,Popa2006Adaptive,Popa2007EKF}, in which the next sample location is optimally computed based on the previous measurements. Such active online sampling is of particular interest in the study of field sensing and reconstruction problems due to its adaptation to the unknown or dynamic field properties. Moreover, from the IoT perspective, they are extremely appealing because many real-time applications require the FC to onlinely deal with a significant amount of streaming sensor data with low latency. Obviously, the aforementioned offline sampling approach is no longer suitable.

 In fact, the idea of active sampling is not totally new. For example, \cite{Willett2004Backcasting} employed a recursive dynamic partition (RDP) based hierarchical approach called ``backcasting'' to reduce communication costs. \cite{Nowak2003Boundary,Rui2005Faster,Rui2008Active} analytically demonstrated that such sensing method achieves faster convergence rate. Yet these works are restricted to the sensing of some specific non-parametric inhomogeneous fields. For more general field sensing problems, mobile robotic sensors were preferred \cite{Suh2016Mobile,Nguyen2015Information,Grasso2016Dynamic,Popa2006Adaptive}. In these works, the key problem lies in how to steer the mobile sensor to the next sensing location in the field based on the information gathered so far. Specifically, for a non-parametric Gaussian process (GP) modeled field, information-theoretic criteria are often utilized to choose the next optimal sensing location. For instance, using a single sensing robot, Suh \emph{et al}. proposed an environmental monitoring navigation strategy, which effectively maximizes the information gain along the robot¡¯s trajectory \cite{Suh2016Mobile}. Considering a team of sensing agents, \cite{Nguyen2015Information} proposed an adaptive sampling strategy that picks out the next location through minimizing the uncertainty, i.e., the conditional entropy at the unobserved locations. As for a parametric field model, Popa \emph{et al}. proposed extended Kalman filter (EKF) \cite{Popa2006Adaptive,Popa2007EKF} based adaptive sampling approach to optimally estimate the parameters.

In summary, online algorithms aim to decline the uncertainty in the knowledge of field distribution. Despite the contributions, these active sampling approaches are faced with challenges on multiple fronts.
First, active sampling inevitably induce complex coordination and frequent communication between sensors \cite{Mysorewala2012A}, thus inducing extra communication cost and computational complexity. Second, robot-like sensors have constrained mobility, which confines the next sensing location to a limited region and degrades the overall convergence rate. Last but not least, the algorithms above are mostly task-specific and cannot be transferred to other field sensing tasks. When the field fluctuates or the task changes, they have to re-execute the entire sensing procedure, which is time-consuming and energy-inefficient. Extending to the IoT paradigm, the current active sampling algorithms incur extra burden in implementation, and may not be able to satisfy the requirements of fast and intelligent ambient sensing.

In this paper, we improve the performance of active sampling algorithms through the subtle intrinsic interaction between communication and computation. Intuitively, communication provides additional data for more accurate computation, and in the meantime, computation has the potential to enable more selective sensing and effective communication along the process. Hence, these two should be exquisitely incorporated to develop an efficient sensing algorithm based on \emph{integrated communication and computation}. In particular, with the help of online reinforcement learning and the state-of-the-art meta-learning techniques, a robust two-layer learning and sensing algorithm, which adaptively determines the most informative sensing location, is presented. It consists of a stochastic-gradient-descent (SGD) based base-learner used for the field model prediction aiming to minimize the average prediction error, and a reinforcement meta-learner aiming to optimize the sensing decision by simultaneously rewarding the error reduction with samples obtained so far and penalizing the corresponding communication cost. It significantly reduces the communication overhead and lays a good foundation for future sensing and fusion system.

To summarize, the contributions of this paper are listed as follows:
\begin{itemize}
	\item A two-layer sensing and learning framework based on meta-learning is proposed for field sensing and reconstruction problems. This two-layer meta-learning framework implies a smart explore-and-exploit strategy, which guides the sensing (exploration) by active learning (exploitation), and in turn improves the learning (exploitation) with effective sensing (exploration).
	\item An adaptive sensing algorithm based on the above two-layer meta-learning framework is presented. It actively determines the most informative sensing location, and thus considerably reduces the spatial samples and yields superior performance and robustness compared with conventional schemes.
	\item The convergence behaviour of the proposed algorithm is also comprehensively analyzed and simulated. The results reveal that for typical scenarios, the proposed field sensing algorithm significantly improves the convergence rate.
\end{itemize}

The rest of the paper is organized as follows. Section
\uppercase\expandafter{\romannumeral2} describes the system model for the specific field sensing and reconstruction problem, and defines
the main objective to be achieved. The adaptive two-layer meta-learning based sensing and learning framework is brought out in Section \uppercase\expandafter{\romannumeral3}, and algorithm design of the meta-learner and the base-learner are also discussed there. The asymptotic performances including the convergence behavior of the proposed framework is analyzed in detail in
Section \uppercase\expandafter{\romannumeral4}. Section
\uppercase\expandafter{\romannumeral5} shows the simulation settings and the comprehensive simulation results.  And finally, Section
\uppercase\expandafter{\romannumeral6} concludes the paper and provides a brief discussion on future works.

\section{System Model and Problem Formulation}
\subsection{System model}
\begin{figure}
\centering
\vspace{0.15in}
\includegraphics[width=0.48\textwidth]{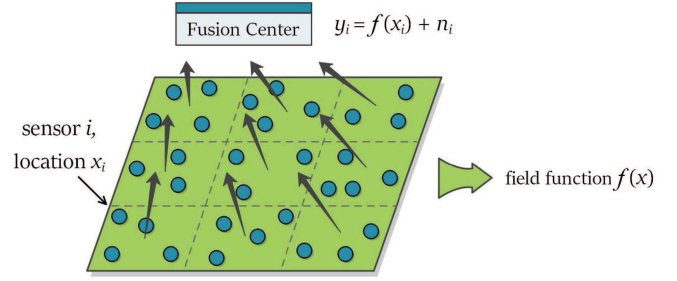}
\caption{Field sensing and reconstruction problem.}
\label{information_c}
\vspace{-0.4cm}
\end{figure}

\begin{figure*}[!hb]
\centering
\includegraphics[width=0.75\textwidth]{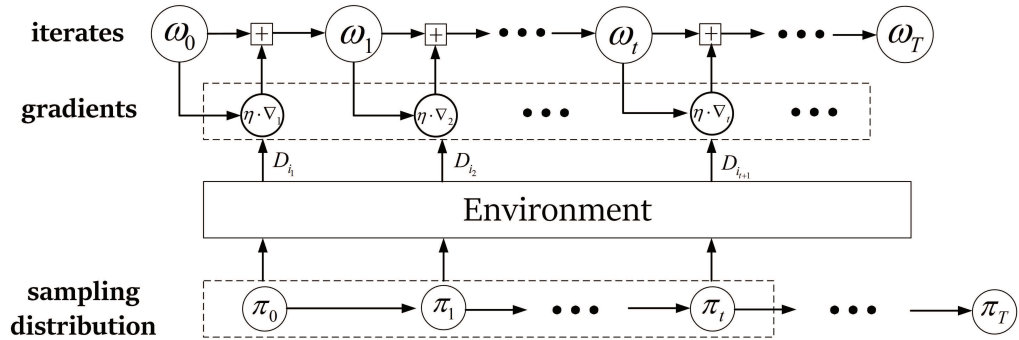}
\caption{$T$-step SGD optimization process.}
\label{sgd}
\end{figure*}

Suppose $n$ sensors are randomly deployed in a $d$-dimensional field to get some scalar quantity which is determined by an unknown field function $f(x): \mathbb{R}^d \rightarrow \mathbb{R}$, as shown in Fig. \ref{information_c}. The noisy measurement at the $i$-th sensor is thus given by
\begin{equation}
y_i=f(x_i)+n_i,
\end{equation}
where $x_i \in \mathbb{R}^d$ is the coordinate of the $i$-th sensor, and $n_i$ is the Gaussian noise. An FC is deployed to collect the measurements from the potential sensors, based on which, the field function $f(x)$ is then reconstructed.

In general, an unknown field can be represented by a combination of parameterized basis (kernel) functions such as the Radial Basis Functions (RBFs) \cite{Buhmann2003RBF} which well captures the local characteristics of almost all nonlinear fields and then effectively approximates the whole fields. Invoking the RBF kernel representation, $f(x)$ can be represented as
\begin{equation}
f(x)=\Phi(x)\boldsymbol{\omega},
\label{fx}
\end{equation}
where $\Phi(x)=\left[\phi_1(x),\phi_2(x),...,\phi_K(x)\right]$ denotes the known RBF kernels and $\boldsymbol{\omega}=\left[\omega_1,\omega_2,...,\omega_K\right]^T$ is the corresponding weight vector. In this paper, for useful insights and ease of treatment, we use the isotropic Gaussian kernel-weighted model and select $\phi_j(x),~j=1,2,...,K,$ as a Gaussian RBF with center $c_j$ and constant width $\beta$. Specifically, we have
\begin{equation}
\phi_j(x)=\textrm{exp}(-\frac{\|x-c_j\|^2}{\beta^2}),
\end{equation}
where $\|x-c_j\|^2$ represents the squared Euclidean distance between sensor location $x$ and the $j$-th kernel center $c_j$. $\beta$ is empirically chosen and characterizes the locally-decaying speed of the spatial phenomenon. Intuitively, a larger $\beta$ potentially leads to a smoother field or vice versa. Given that $\Phi(x)$ is known and fixed, now the FC only needs to find a good estimation of the weight vector $\boldsymbol{\omega}$.



\subsection{Field Reconstruction Based on Sensed Samples}

Given enough sensed data, the optimal $\boldsymbol{\omega}$ can be obtained by solving the following optimization problem which minimizes the overall loss function:
\begin{equation}
\underset{\boldsymbol{\omega}\in \mathbb{R}^K}{\min}\textrm{P}(\boldsymbol{\omega}):=\underbrace{\frac{1}{n}\sum_{i=1}^{n}L_{i}(\boldsymbol{\omega})}_{L(\boldsymbol{\omega})}+\gamma\Gamma(\boldsymbol{\omega}),
\label{goal1}
\end{equation}
where $L_i(\boldsymbol{\omega})=\left(\Phi(x_i)\boldsymbol{\omega}-y_i\right)^2$ is the square error at location $x_i$ w.r.t. $\boldsymbol{\omega}$,  $L(\boldsymbol{\omega})$ is the average loss over $n$ training data samples $\{D_{1},...,D_{n}\}$ with $D_{i}\triangleq (x_i,y_i)$, and $\Gamma(\boldsymbol{\omega})=\|\boldsymbol{\omega}\|_1$ is the regularizer which ensures the sparsity of $\boldsymbol{\omega}$, and $\gamma > 0$ is the regularization parameter.

As a typical variant of \emph{stochastic approximation} \cite{Robbins1951A}, the above L1-norm regularized optimization problem \eqref{goal1} can be recursively solved by a well known method called proximal gradient descent, which can be described by the following update rule for $t=1,2,...$ \cite{Zhao2015Stochastic}:

\begin{equation}
\begin{split}
&\boldsymbol{\omega}_0 \in \mathbb{R}^K, \\
&\boldsymbol{\omega}_{t+1} =\textrm{prox}_{\gamma \eta_t\Gamma}\left(\boldsymbol{\omega}_t - \eta_t \hat{\xi}_t(\boldsymbol{\omega}_t) \right),
\end{split}
\label{recursive}
\end{equation}
where $\boldsymbol{\omega}_0$ is the initial weight vector. Defining the proximal operator as $\textrm{prox}_h(x)=\arg \min_{\boldsymbol{\omega}}h(\boldsymbol{\omega})+\frac{1}{2}\|\boldsymbol{\omega}-x\|^2$
, the above rule means that the  $t+1$-th iterate satisfies:
\begin{equation}
\boldsymbol{\omega}_{t+1}=\arg \min_{\boldsymbol{\omega}} \gamma\eta_t\Gamma({\boldsymbol{\omega}})+\frac{1}{2}\|\boldsymbol{\omega}-\left(\boldsymbol{\omega}_{t}-\eta_t \hat{\xi}_t(\boldsymbol{\omega}_t) \right)\|^2,
 \end{equation}
 where  $\hat{\xi}_t$ denotes an estimate of the gradient $\nabla L(\boldsymbol{\omega}_t)$ at Step $t$, and $\eta_t$ is the learning rate.
Further, given $\Gamma({\boldsymbol{\omega}})=\|{\boldsymbol{\omega}}\|_1$, the proximal mapping is the following shrinkage operation, also known as the  ``soft-threshold operator'':
\begin{equation}
\textrm{prox}_{\gamma \eta_t \Gamma}(\boldsymbol{\omega})=\textrm{sign}(\boldsymbol{\omega})\odot[|\boldsymbol{\omega}|-\gamma \eta_t]_{+},
\end{equation}
with $[\cdot]_{+} \triangleq \max(\cdot,0)$.

To calculate $\hat{\xi}_t$, one commonly appeals to batch gradient descent (BGD) or stochastic gradient descent (SGD), as in traditional statistical learning approaches. Since BGD requires the whole dataset to estimate the gradients (i.e., $\hat{\xi}_t(\boldsymbol{\omega}_t)=\nabla L(\boldsymbol{\omega}_t)$ for Step $t\geq 0$), while SGD only uses the data sensed by a randomly selected sensor (i.e., $\hat{\xi}_t(\boldsymbol{\omega}_t)=\nabla L_{i_{t+1}}(\boldsymbol{\omega}_t)$ for Step $t+1$), it is generally less efficient for BGD to be applied to the studied scenario considering its larger overall communication cost and delay needed for the reconstruction. As such, we mainly focus on the SGD approach in this paper, which is illustrated in Fig. \ref{sgd}. Specifically, the environment corresponds to the field which generates data samples at each sensor. In each step, after a new data is sensed, the gradients are calculated based on the previous weight vector and are then used to produce the next one.

\subsection{How to sense efficiently: uniform or non-uniform sampling?}

Due to the largely unknown parameters of the field, reconstructing $f(x)$ from a totally randomly selected data (like the vanilla SGD) at the FC in general still requires a large dataset and thus consumes plenty of communication and computing resources, which makes the field sensing and reconstruction costly and practically inefficient. One intuitive way to relieve this situation is to explore and exploit the most informative data samples from all potential sensors based on the already observed samples, i.e., the field sensing is driven by certain kinds of statistical learning and prediction so as to reduce the overall cost of sensing and communication.

In other words, to enable fast and accurate reconstruction, it is rather crucial to do efficient selective sensing / sampling, i.e., the sampling distribution in the $T$-step SGD process as illustrated in Fig. \ref{sgd} should be carefully designed. Note that for vanilla SGD, the sampling distribution for Step $t$, $\boldsymbol{\pi}_t=[p_1^{(t)},\cdots,p_n^{(t)}]$ are in fact constant, or namely, $p_i^{(t)} = 1/n$. By using such a uniform sampling, on one hand, fast startup could be achieved since it does not need all the data to be ready in advance. On the other hand, only single derivative is  calculated thus the per iterate computational cost is reduced to $1/n$ in comparison with BGD. However, imaginably, the purely random sampling also inevitably introduces larger data redundancy due to the intrinsic spatial correlation of the field, as well as poorer convergence due to the large deviation of $\nabla L_i (\boldsymbol{\omega}_t)$ with the index $i$.

One recent promising approach to improve the SGD performance is to incorporate adaptive non-uniform sampling with it. As shown in many recent studies (see \cite{Zhao2015Stochastic, Needell2016Stochastic} and references therein), sampling at a probability distribution in proportion to the relative importance of a data sample with respect to the entire dataset, as represented by its relative norm
\begin{equation}
p_i^{t}=\frac{\|\nabla L_i(\boldsymbol{\omega}_t)\|}{\sum_{j=1}^{n}  \|\nabla L_j(\boldsymbol{\omega}_t)\|},
\label{importance-sampling}
\end{equation}
can achieve certain optimum in terms of efficiency and prediction error. Unfortunately, such importance-based sampling has to evaluate the gradients based on the entire dataset and relies on the full knowledge of the target model, which is largely unknown until the sensors are really chosen to sense and send its data to the FC.

\begin{figure}[!ht]
\centering
\includegraphics[width=0.48\textwidth]{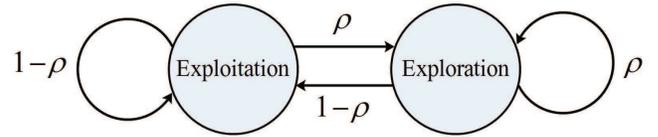}
\caption{Markov chain-based sampling system.}
\label{exp_lor_loi}
\vspace{-0.3cm}
\end{figure}

Therefore, it is desirable to develop some adaptive sampling algorithm which is able to exploit the information incrementally gathered so far while keeping the capability to explore the unknown portion of the field as the algorithm proceeds. Motivated by this, we enforce the following Markovian greedy sampling scheme:
\begin{equation}
\boldsymbol{\pi}_t=\rho \boldsymbol{\pi}_{u}+ (1-\rho)\boldsymbol{\pi}_{v,t},
\label{markovpi}
\end{equation}
where $\boldsymbol{\pi}_{u}$ denotes the fixed uniform sampling distribution as used in vanilla SGD, $\boldsymbol{\pi}_{v,t}$ is the importance-based sampling distribution varying with $t$ as defined in \eqref{importance-sampling} and calculated over the data sampled so far, i.e., $D_{i_1},...D_{i_t}$, and $\rho \in [0,1]$ is a tuning parameter. The rationale behind the above equation can be interpreted as follows:
the resultant sampling distribution $\boldsymbol{\pi}_t$ is a mixture of two laws of distribution that stand for two complementary tendencies of sampling strategy, respectively. The former tends to fully \emph{explore} the unknown field while the latter tends to effectively \emph{exploit} the gathered information so far. As such, the above sampling process can be regarded to work in two different states, referred to as ``\emph{exploration}'' and ``\emph{exploitation}'', respectively, as depicted in the Markov chain in Fig.\ref{exp_lor_loi}.

\section{Two-layer Learning and Sensing Algorithm}

The Markov chain based sampling framework in the previous Section realizes some trade-off between exploration and exploitation, yet it is not smart enough to avoid all redundant or less significant samples since when the weighting factor $\rho$ is fixed, the uniform random sampling always exists even if the amount of samples are large enough and the convergence is approached.

As a remedy, we can further use a time-varying weight $\rho_t$ to tune the proportion of distributions for more flexible trade-off between exploration and exploitation. To this end, we resort to the recently emerged idea in the generic area of machine learning, a.k.a. ``learn to learn'' or ``meta-learning'' \cite{Andrychowicz2016Learning,Ravi2017Optimization,Li2016Learning,Li2017Learning} which enables a learning machine to learn its own learning process so as to learn more intelligently, and propose the so-called ``learn to sense'' framework which enables a smarter online adjustment of sensing strategy and faster field reconstruction with much less sensing / communication effort.

\subsection{Two-layer learning and sensing framework}

In particular, we propose a two-layer learning and sensing framework, which includes a conventional SGD-based base learner and a high-level reinforcement-learning-based meta-learner, as shown in Fig. \ref{reinforce-base-meta-learner}. The meta-learner aims to generate an optimal sampling policy $\boldsymbol{\pi}_t$ that minimizes the ``meta-loss'' (or equivalently, maximizes the overall reward) which serves as an integrated measure of both the processing gain of the base learner and the sensing / communication cost of the environment, which is defined as follows:
\begin{equation}
L_{\textrm{meta}}=[L(\boldsymbol{\omega}_T)-L(\boldsymbol{\omega}_0)]+\mu |O_T|.
\label{meta-loss}
\end{equation}
In the above equation, the first term in the brackets quantifies the increment of average loss (prediction error) at Step $T$ w.r.t. the time of start, while the second term denotes the overall sensing / communication cost paid for the sampled dataset $O_T$ with a unit price of $\mu$. Note that the first term is in general negative and tends smaller as the algorithm proceeds.

In the above two-layer meta-learning framework, the iterative interaction between the meta-learner and the base learner is crucial to obtain the optimal sampling policy as well as the desired reconstruction performance. A reinforcement learning algorithm based on partially-observable Markov decision process (POMDP) is employed for this purpose with details described below.

\subsection{Meta-learner and base-learner design}

\begin{figure}[!t]
\centering
\includegraphics[width=0.48\textwidth]{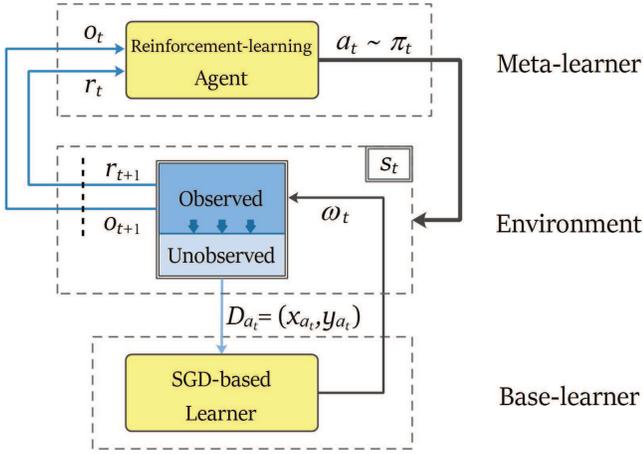}
\caption{The two-layer learner based on reinforcement learning.}
\label{reinforce-base-meta-learner}
\vspace{-0.3cm}
\end{figure}

The basic algorithms of the meta-learner and the base-learner are designed in the following. As shown in Fig. \ref{reinforce-base-meta-learner}, this two-layer learner can be further transformed into a POMDP-based policy training machine. The basic factors of the related tuple $<s_t,o_t,a_t,\boldsymbol{\pi}_t,r_t>$ are defined in detail as follows:
\begin{itemize}
	\item $s_t$ is the state represented by the entire dataset and the corresponding status of each data at time $t$ indicating whether it is  observed or not. Note that the fusion center makes incremental observation to the field and thus only part of the data are known to it at each time. 
	\item $o_t$ contains the set of observed data at time $t$.
	\item $a_t$ denotes the action made by the meta-learner at time $t$ and here is the index of the next sensor location to be sampled.
	\item $\boldsymbol{\pi}_t$ stands for the sampling policy at time $t$ and here is the conditional probability distribution over the action space $\{1,2,...,n\}$ given the current observed measurements $o_t$.
	\item $r_t=R(s_t,a_t)$ denotes the reward value that the previous action gains at the current state. Invoking \eqref{meta-loss}, here we let $R(s_t,a_t)=\mbox{Loss Gap} - \textbf{1}(a_t \notin O_{t-1})\times \mu$, where ``\mbox{Loss Gap}'' is the prediction error reduced by the model update, $\mu$ is the unit cost induced by extra sensing / communication, and $O_t$ is the set of sensor indices corresponding to $o_t$.
\end{itemize}

The system works in an iterative manner. In iteration $t$, the agent activates the $a_t$-th sensor to get its data sample $D_{a_t}=(x_{a_t},y_{a_t})$ according to the current sampling policy $\boldsymbol{\pi}_t$. The sampled data is then passed to the base-learner which makes the prediction based on SGD and results in an updated model $\boldsymbol{\omega}_t$. With the set of sensed data or the field information partially observed by the agent grown from $o_t$ to $o_{t+1}$, a state transition occurs, which triggers an update of the reward to $r_{t+1}=R(s_t,a_t)$. With the new observation $o_{t+1}$ and reward $r_{t+1}$, the agent generates a new sampling policy $\boldsymbol{\pi}_{t+1}$ for the next iteration.

\begin{figure}
\centering
\includegraphics[width=0.45\textwidth]{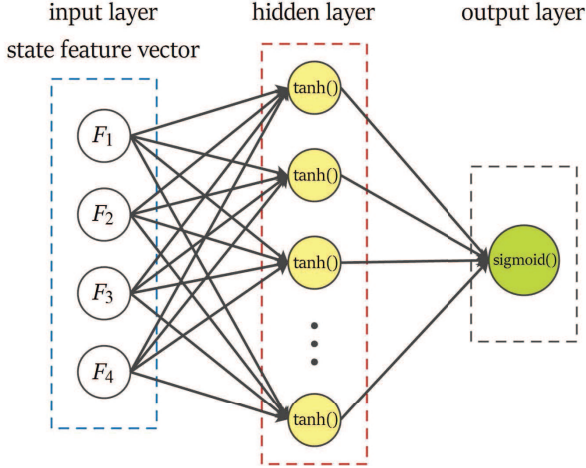}
\caption{The three-layer neural policy network.}
\label{policynn}
\end{figure}

Now let us elaborate more on the meta-learner which tries to learn the optimal sampling policy $\boldsymbol{\pi}_{t}^*$ with a goal maximizing the expected reward accumulated over time, i.e.,
\begin{equation}
\boldsymbol{\pi}_t^* = \arg \max_{\boldsymbol{\pi}_t} E\Big[\sum_{t=0}^{T}R(s_t,a_t)\Big],
\label{pi_opt}
\end{equation}
in which the expectation is taken over the sequence of states and actions $\{s_0,a_0,s_1,a_1,...,s_T\}$. And for effectiveness and learnability, $\boldsymbol{\pi}_t$ is often supposed to be within a pmf family parameterized by $\Theta$ as follows:
\begin{align}
\boldsymbol{\pi}_t(\Theta)
&=P_{\Theta}\left(a_t|o_t\right) \notag\\
&=P_f(a_t|o_t) \rho_t(o_t,\Theta)+P_v(a_t|o_t) \left(1-\rho_t(o_t,\Theta)\right),
\label{pi_Theta}
\end{align}
where $P_f(a_t|o_t)\equiv \frac{1}{n}$ refers to the fixed uniform sampling distribution over the whole dataset, while $P_v(a_t=i|o_t)$ is the  importance sampling distribution associated with the dataset varying over time and is defined as follows:
\begin{equation}
P_v(a_t=i|o_t)=\frac{g_i(t)}{\sum_{i=1}^{n}g_i(t)},
\end{equation}
where
\begin{equation}
g_i(t)=
\begin{cases}
\|\nabla L_i(\boldsymbol{\omega}_t)\| & i \in O_t \\
0 & \text{otherwise}
\end{cases}.
\end{equation}
Moreover, $\rho_t(o_t,\Theta)$ is the dynamic weight varying as the learning proceeds. From (\ref{pi_Theta}) we can see that the optimal $\Theta$, i.e., $\Theta^*$, can be solely determined from $\rho_t(o_t,\Theta)$. Therefore, the goal reduces to finding the best parameterized non-linear mapping from $o_t$ to $\rho_t\ \in [0,1]$. For simplicity, a three-layer neural network \cite{Li2017Learning} is chosen to model this non-linear mapping, but note that other deep neural works can also be used. As shown in Fig. \ref{policynn}, the input layer of the three-layer neural network includes the following features:
\begin{align}
&F_1=\textbf{1}(a_{t-1} \in O_{t-1}),\\
&F_2=\frac{L_{O_{t-1}}(\omega_t)-L_{O_{t-1}}(\omega_{t-1})}{L_{O_{t-1}}(\omega_{t-1})},\\
&F_3=\frac{L_{O_t}(\omega_t)-L_{O_{t-1}}(\omega_{t-1})}{L_{O_{t-1}}(\omega_{t-1})},\\
&F_4=\frac{|O_t|L_{O_t}(\omega_t)-|O_{t-1}|L_{O_{t-1}}(\omega_{t-1})}{|O_{t-1}|L_{O_{t-1}}(\omega_{t-1})},
\end{align}
In which $L_{O_t}(\boldsymbol{\omega}_t)$ denotes the average loss associated with the already sampled sensors in $O_t$ and the current model $\boldsymbol{\omega}_t$. In particular, $F_1$ indicates whether $D_{t-1}$ is directly sampled from the existing observed dataset or not, $F_2$ shows how much $D_{t-1}$ weighs on the model estimation in the previous step, whereas $F_3$ indicates how much it does on the current step, and finally $F_4$ indicates the relative loss change after the model update. The hidden units in the middle layer are activated by $\textrm{tanh}(\cdot)$, whereas $\textrm{sigmoid}(\cdot)$ are used in the output layer to ensure $\rho_t(o_t,\Theta^*)\in (0,1]$.

Based on \eqref{pi_Theta}, the original policy optimization problem \eqref{pi_opt} reduces to
\begin{equation}
\max_{\Theta} J(\Theta)=\max_{\Theta} E_{P_{\Theta}(a|s)} R(s,a),
\label{J}
\end{equation}
where $R(s,a)$ is the state-action value function.
Although $R(s,a)$ is non-differentiable w.r.t. $\Theta$,
\begin{equation}
\begin{aligned}
\nabla J(\Theta) &= \sum_{\tau}R(\tau)P(\tau|\Theta) \frac{\nabla P(\tau|\Theta)}{P(\tau|\Theta)} \\
&=\sum_{\tau}R(\tau)P(\tau|\Theta)\nabla\log P(\tau|\Theta),
\end{aligned}
\end{equation}
where an episode is considered as a trajectory $\tau=\{s_1,a_1,r_1,...,s_T,a_T,r_T\}$, and $R(\tau)$ is the cumulative reward obtained from one episode. By summing over the expected rewards at all $T$ steps, the above equation can be rewritten as:
\begin{equation}
\nabla_{\Theta} J=\sum_{t=1}^{T} E_{P_{\Theta}(a_{1:T}|s)} \Big[R(s_t,a_t)\nabla_{\Theta}\log P_{\Theta}(a_t|s_t)\Big],
\label{monte}
\end{equation}
Invoking the famous Monte-Carlo policy gradient algorithm REINFORCE \cite{Williams1992Simple} (see Algorithm 1), the above can be empirically estimated as:
\begin{equation}
\sum_{t=1}^{T} \nabla_{\theta} \log P(a_t|s_t)v_t.
\end{equation}
Here $v_t$ is the sampled estimation of $R(s_t,a_t)$ from one episode execution  of the current sampling policy $P_\Theta(a|s)$: $v_t = r_t + \lambda r_{t+1} + ...+\lambda^{T-t} r_T$, in which $r_t$ is the instantaneous reward at step $t$ and $\lambda \in [0,1]$ is the discount factor.
\begin{algorithm}[H]
\caption{REINFORCE\cite{Williams1992Simple}}
\begin{algorithmic}
\STATE Initialize $\theta$ with arbitrary value
\FOR {each episode $\{s_{1},a_{1},r_{2},\cdots,s_{T-1},a_{T-1},r_{T}\}$}
	\FOR {$t=1$ to $T-1$}
		\STATE $\theta\leftarrow \theta + \alpha \nabla_{\theta} \log \pi_{\theta} (s_{t},a_{t})v_{t}$
	\ENDFOR
\ENDFOR
\RETURN $\theta$
\end{algorithmic}
\end{algorithm}

\section{Asymptotic Performance Analysis}

In this section, we provide an asymptotic behavior and performance analysis of the proposed algorithm.

\subsection{Preliminaries}
Here, we briefly introduce some key definitions and propositions that are useful there-in-after.

\emph{Definition 1:} A function $\phi: \mathbb{R}^{d}\rightarrow \mathbb{R}$ is $L$-Lipschitz if for all $u,~v\in \mathbb{R}^{d}$ we have
\begin{equation}
\|\phi(u)-\phi(v)\|\leq L\|u-v\|,
\end{equation}
where $\|\cdot\|$ is a norm.

\emph{Definition 2:} A function $\phi: \mathbb{R}^{d}\rightarrow \mathbb{R}$ is $(1/\gamma)$-smooth if it is differentiable and its gradient is $(1/\gamma)$-Lipschitz.

\emph{Definition 3:} A function $\phi: \mathbb{R}^{d}\rightarrow \mathbb{R}$ is $\sigma$-strongly convex if for all $u,~v\in \mathbb{R}^{d}$ we have
\begin{equation}
\phi(u)\geq \phi(v)+\nabla\phi^{T}(v)(u-v)+\frac{\sigma}{2}\|u-v\|^{2}.
\end{equation}
%
%
Based upon the above definitions, we give the following lemma.

\vspace{0.1cm}
\textbf{\emph{Lemma 1:}} i) $L_{i}(\boldsymbol{\omega})$ is $(2n)$-smooth for $i=1,2,...,n$, ii) $L(\boldsymbol{\omega})=\frac{1}{n}\sum_{i=1}^{n}L_{i}(\boldsymbol{\omega})$ is $2$-strongly convex.

\begin{IEEEproof}
Recall that $L_{i}(\boldsymbol{\omega})=\left(\Phi(x_i)\boldsymbol{\omega} - y_i\right)^2$, thus the gradient equals
\begin{equation}
\nabla L_{i}(\boldsymbol{\omega})=-2\left(\Phi(x_i)\boldsymbol{\omega} - y_i\right)\Phi^{T}(x_i).
\end{equation}
Due to the Gaussian kernel-based model assumption, we have $\|\Phi(x_i)\|^2\leq n^2$. Then for all $\boldsymbol{\omega},~\boldsymbol{\omega}'\in \mathbb{R}^{K}$, the following inequality holds.
\begin{align}
\|\nabla L_{i}(\boldsymbol{\omega})-\nabla L_{i}(\boldsymbol{\omega}')\|
&=2\|\Phi(x_i)(\boldsymbol{\omega}-\boldsymbol{\omega}')\|\|\Phi^{T}(x_i)\|\notag\\
&\leq 2\|\Phi(x_i)\|\|\boldsymbol{\omega}-\boldsymbol{\omega}'\|\|\Phi^{T}(x_i)\| \notag\\
&=2\|\Phi(x_i)\|^{2}\|\boldsymbol{\omega}-\boldsymbol{\omega}'\| \notag\\
&\leq 2n \|\boldsymbol{\omega}-\boldsymbol{\omega}'\|.
\end{align}
Obviously, the gradient $\nabla L_{i}(\boldsymbol{\omega})$ is $(2n)$-Lipschitz constant, therefore $L_{i}(\boldsymbol{\omega})$ is $(2n)$-smooth.

To show $L(\boldsymbol{\omega})$ is $\sigma$-strongly convex, it is necessary to prove the following inequality for all $\boldsymbol{\omega},~\boldsymbol{\omega}'\in \mathbb{R}^{K}$.
\begin{equation}
\label{lemma1neces}
L(\boldsymbol{\omega})\geq L(\boldsymbol{\omega}')+\nabla L^{T}(\boldsymbol{\omega}')(\boldsymbol{\omega}-\boldsymbol{\omega}')+\|\boldsymbol{\omega} - \boldsymbol{\omega}'\|^{2}.
\end{equation}

With subsitution $\tilde{\boldsymbol{\omega}}:=y_i-\Phi(x_i)\boldsymbol{\omega}$, $\tilde{\boldsymbol{\omega}}':=y_i-\Phi(x_i)\boldsymbol{\omega}'$. and by setting $\sigma=2$, we have
\begin{align}
&L_{i}(\boldsymbol{\omega}')+\nabla L_{i}(\boldsymbol{\omega}')^{T}(\boldsymbol{\omega}-\boldsymbol{\omega}') + \|\boldsymbol{\omega} - \boldsymbol{\omega}'\|^{2} \notag\\
&=\tilde{\boldsymbol{\omega}}'^{2}+2\tilde{\boldsymbol{\omega}}'(\tilde{\boldsymbol{\omega}} - \tilde{\boldsymbol{\omega}}') + \frac{1}{\|\Phi(x_i)\|^{2}}\|\tilde{\boldsymbol{\omega}} - \tilde{\boldsymbol{\omega}}'\|^{2}\notag\\
&\leq \tilde{\boldsymbol{\omega}}'^{2}+2\tilde{\boldsymbol{\omega}}'(\tilde{\boldsymbol{\omega}} - \tilde{\boldsymbol{\omega}}') + \|\tilde{\boldsymbol{\omega}} - \tilde{\boldsymbol{\omega}}'\|^{2} \notag\\
&=\tilde{\boldsymbol{\omega}}^{2} = L_{i}(\boldsymbol{\omega}).
\end{align}




Summing over $i=1,\cdots,n$, and invoking $L(\boldsymbol{\omega})=\frac{1}{n}\sum_{i=1}^{n}L_{i}(\boldsymbol{\omega})$ directly yields \eqref{lemma1neces} and completes the proof.
\end{IEEEproof}

\subsection{Main results}
Now we now turn to the analysis of the asymptotic performances of our proposed sensing algorithm.

Define $\boldsymbol{\omega}^*$ the optimal solution, and introduce the sampling probability
\begin{equation}
\begin{aligned}
\boldsymbol{\pi}^*&=(\pi_1^*,...,\pi_N^*) \\
     &=\boldsymbol{\pi}_{u} \rho^*+\boldsymbol{\bar{\pi}}^*(1-\rho^*),
\end{aligned}
\end{equation}
where $\boldsymbol{\bar{\pi}}^*$ is the ideal sampling distribution defined by \eqref{importance-sampling}, and $\rho^*$ is the optimal weight parameter learned by our policy training scheme.

\vspace{0.1cm}
\textbf{\emph{Lemma 2:}} Denote $Q^*=\sum_{i=1}^{N} \nabla L_i(\boldsymbol{\omega}^*) \nabla L_i(\boldsymbol{\omega}^*)^{T}/(n^2 \bar{p}_i^*)$ and $H=\nabla^2 L(\boldsymbol{\omega}^*)$ the Hessian at point $\boldsymbol{\omega}^*$, and set $\eta_{t}=\frac{1}{2t}$, then we have:
\begin{enumerate}
   \item The sequence $(\boldsymbol{\omega}-\boldsymbol{\omega}^*)/\sqrt{\eta_t}$ converges to a zero-mean Gaussian variable $V \sim N(0,\Sigma)$, where the covariance matrix $\Sigma$ is the solution to the following Lyapunov equation
\begin{equation}
\Sigma(I_K+H)+(I_K+H)\Sigma=Q^*,
\end{equation}
   \item The sequence $\left(L(\boldsymbol{\omega})-L(\boldsymbol{\omega}^*)\right)/\sqrt{\eta_t}$ converges to a random variable $V'=(1/2)Z^T\Sigma^{1/2}H\Sigma^{1/2}Z$ where $Z$ is a Gaussian vector $N(0,I_K)$. The mean of $V'$ is $\mathbb{E}(V')=tr(H\Sigma)/2$.
 \end{enumerate}
\begin{IEEEproof}
Note that the above lemma is the direct result of \cite{Papa2015Adaptive}, by utilizing Lemma 1 and the second-order delta-method. Detailed proof is omitted here due to lack of space.
\end{IEEEproof}

Intuitively, the optimal fixed importance-based sampling distribution in \cite{Zhao2015Stochastic}, should minimize the mean value $\mathbb{E}(V')$, that is,
\begin{equation}
\arg \min_{\boldsymbol{\pi}} tr(H\Sigma) = \boldsymbol{\bar{\pi}}^*
\end{equation}
Set ${v^*}^2=tr(H\Sigma)|_{{\boldsymbol{\pi}}=\boldsymbol{\bar{\pi}}^*}$, the following theorem directly follows from simple standard algebra and some necessary reorganization.

\vspace{0.1cm}
\textbf{\emph{Theorem 1:}} Let $\Sigma$ be the asymptotic covariance matrix defined in Lemma 2. Then,
\begin{equation}
{v^*}^2 \leq tr(H\Sigma) \leq \frac{{v^*}^2}{(1-\rho^*)}
\end{equation}

\emph{Remark:} Theorem 1 implies that normalized error sequence is strictly bounded and more importantly, the asymptotic performance of the proposed algorithm is comparable with the one associated with the best sampling distribution, provided that $\rho^*$ is close enough to zero, which however, is not supposed to happen in the early stage of the online algorithm, since the system needs a non-zero $\rho$ to explore the field. Whereas as $t \to \infty$, consider all the samples have been collected, then $\rho$ can be ideally set to zero, which makes $\boldsymbol{\pi}_t \to \boldsymbol{\bar{\pi}}^* $ with probability 1.

In this sense, as \eqref{markovpi} being the mixture of the uniform and the non-uniform sampling laws, it is safe to argue that its performance falls in between. The lower bound is associated with the uniform sampling vanilla SGD, and has already shown a convergence rate of $O(1/\sqrt{t})$\cite{Zhang2004Solving}. The upper bound, whereas, corresponds to the non-uniform sampling defined by (\ref{importance-sampling}), whose optimality is shown as below.

\vspace{0.1cm}
\textbf{\emph{Theorem 2:}} When the non-uniform sampling probability satisfies (\ref{importance-sampling}), it maximizes the reduction of the objective value defined by the RHS of (\ref{goal1}).

\begin{IEEEproof}
Recall that for proximal SGD with non-uniform sampling, where $p_{i}^{t}$ denotes the sampling probability of the $i_t$-th sample at step $t$, the update rule is written as:
\begin{equation}
\begin{aligned}
&\boldsymbol{\omega}_{t+1}=\\
&\arg \min_{\boldsymbol{\omega}} \left[\langle(n p_{i}^{t})^{-1} \nabla L_{i}(\boldsymbol{\omega}_{t}),\boldsymbol{\omega} \rangle+\gamma\Gamma(\boldsymbol{\omega})
+\frac{1}{2\eta_{t}}\|\boldsymbol{\omega}-\boldsymbol{\omega}_{t}\|^2 \right]
\end{aligned}
\end{equation}
By setting the derivative of optimization function above as zero, we can easily obtain the following implicit solution:
\begin{equation}
\boldsymbol{\omega}_{t+1}=\boldsymbol{\omega}_{t}-\eta_{t}(n p_{i}^{t})^{-1}\nabla L_{i}(\boldsymbol{\omega}_{t})-\eta_{t}\gamma\cdot \partial \Gamma(\boldsymbol{\omega}_{t+1}).
\end{equation}
Invoking Lemma 1, and setting $\eta_{t} \leq 1/{2n}$, we have
\begin{align}\notag
L(\boldsymbol{\omega}_{t+1}) \leq &L(\boldsymbol{\omega}_{t})-\eta_{t} \langle \nabla L(\boldsymbol{\omega}_{t}),\nabla L_{i}(\boldsymbol{\omega}_{t})+\gamma\partial\Gamma(\boldsymbol{\omega}_{t+1}) \rangle \notag\\
&+ n \eta_{t}^2 \|(n p_{i}^{t})^{-1}\nabla L_{i}(\boldsymbol{\omega}_{t})+\gamma\partial\Gamma(\boldsymbol{\omega}_{t+1})\|^2 \notag\\
\leq &L(\boldsymbol{\omega}_{t})-\eta_{t} \langle \nabla L(\boldsymbol{\omega}_{t}),\nabla L_{i}(\boldsymbol{\omega}_{t})+\gamma\partial\Gamma(\boldsymbol{\omega}_{t+1}) \rangle \notag\\
&+ \frac{ \eta_{t}}{2} \|(n p_{i}^{t})^{-1}\nabla L_{i}(\boldsymbol{\omega}_{t})+\gamma\partial\Gamma(\boldsymbol{\omega}_{t+1})\|^2
\end{align}
In addition, since $\Gamma$ is convex,
\begin{align}
&\Gamma(\boldsymbol{\omega}_{t+1}) \leq\Gamma(\boldsymbol{\omega}_{t})+\langle \partial \Gamma (\boldsymbol{\omega}_{t+1}), \boldsymbol{\omega}_{t+1}-\boldsymbol{\omega}_{t}\rangle \notag\\
&=\Gamma(\boldsymbol{\omega}_{t})-\eta_t\langle\partial\Gamma(\boldsymbol{\omega}_{t+1}), (n p_{i}^{t})^{-1}\nabla L_{i}(\boldsymbol{\omega}_{t})+\gamma\partial\Gamma(\boldsymbol{\omega}_{t+1})\rangle.
\end{align}
Combining the above two inequalities, we can get the reduction on the objective function, bounded as:
\begin{align}
&\mathbb{E}P(\omega_{t+1})\notag\\
=&\mathbb{E}\left[L(\omega_{t+1})+\gamma\Gamma(\boldsymbol{\omega}_{t+1})\right] \notag\\
\leq& \mathbb{E}\left[L(\omega_{t})+\gamma\Gamma(\boldsymbol{\omega}_{t})\right]\notag\\
&-\eta_{t}\mathbb{E}\langle \nabla L(\boldsymbol{\omega_t}),\nabla L_{i}(\boldsymbol{\omega}_{t})+\gamma \partial \Gamma(\boldsymbol{\omega}_{t+1})\rangle \notag\\
&+\mathbb{E}\frac{\eta_{t}}{2}\|(n p_{i}^{t})^{-1}\nabla L_{i}(\boldsymbol{\omega}_{t})+\gamma \partial \Gamma(\boldsymbol{\omega}_{t+1})\|^2\notag\\
&-\mathbb{E}\eta_{t}\langle\gamma \partial \Gamma(\boldsymbol{\omega}_{t+1}), (n p_{i}^{t})^{-1}\nabla L_{i}(\boldsymbol{\omega}_{t})+\gamma\partial\Gamma(\boldsymbol{\omega}_{t+1})\rangle \notag\\
=&\mathbb{E}P(\boldsymbol{\omega}_{t})-\frac{\eta_t}{2}\mathbb{E}\|\nabla L(\boldsymbol{\omega}_{t})+\gamma \partial \Gamma(\boldsymbol{\omega}_{t+1}\|^2\notag\\
&+\frac{\eta_t}{2}\sum_{i=1}^{n}p_i^t\| (n p_{i}^{t})^{-1}\nabla L_i(\boldsymbol{\omega}_t-\nabla
L(\boldsymbol{\omega}_t)\|^2 \notag\\
=&\mathbb{E}P(\boldsymbol{\omega}_{t})-\frac{\eta_t}{2}\mathbb{E}\|\nabla L(\boldsymbol{\omega}_{t})+\gamma \partial \Gamma(\boldsymbol{\omega}_{t+1}\|^2\notag\\
&+\frac{\eta_t}{2n^2}\sum_{i=1}^{n}(p_i^t)^{-1}\|\nabla L_{i}(\boldsymbol{\omega}_t)\|^2-\frac{\eta_t}{2}\|\nabla L(\boldsymbol{\omega}_t)\|^2.
\label{lbound}
\end{align}
Therefore, in order to maximize the reduction on the objective value, it is straightforward that the third term in \eqref{lbound} should be minimized, so that the ideal choice of $\{p_{i}^{t}\}$ turns out to be
\begin{equation}
\label{ideal}
p_i^{t*}=\frac{\|\nabla L_i(\boldsymbol{\omega}_t)\|}{\sum_{j=1}^{n} \|\nabla L_j(\boldsymbol{\omega}_t)\|}.
\end{equation}
\end{IEEEproof}

Although the above optimality can not be achieved immediately, it can be gradually approximated using the accumulated samples in our proposed scheme.

\vspace{0.1cm}
\textbf{\emph{Corollary 1:}} The proposed meta-learning based algorithm performs better if samples located at steep slopes of the original field are collected sooner.

\begin{IEEEproof}
First, we write
\begin{equation}
\|\nabla L_{i}(\boldsymbol{\omega}_{t})\|=2|y_{i}-\Phi(x_{i})\boldsymbol{\omega}_t|\cdot \|\Phi(x_i)\|.
\end{equation}
Since $\|\phi(x_i)\|\approx\|\phi(x_j)\|$ for all $i,j\in 1,...,n$,
\eqref{ideal} can be approximated by:
\begin{equation}
p_i^{t*}\approx\frac{|y_{i}-\Phi(x_{i})\boldsymbol{\omega}_t|}{\sum_{j=1}^{n} |y_{j}-\Phi(x_{j})\boldsymbol{\omega}_t|}
\end{equation}
Therefore, it is straightforward that the ideal sampling distribution is approximately determined by the distribution of residual error, $E^t=|y-\Phi(x)\boldsymbol{\omega}_t|$ over the entire field. By taking derivative of $E^t$ w.r.t $x$, we have
\begin{equation}
\frac{d E^t}{d x}=\left|\frac{dy}{dx}-\frac{d\Phi(x)}{dx}\boldsymbol{\omega}_t\right|=\left|\frac{dy}{dx}-\frac{d \tilde{y}^t}{dx}\right|,
\end{equation}
where $\tilde{y}^t$ stands for the field estimation at step $t$.

The above equation reveals that the the shape of error function at step $t$, is well captured by the difference between the first-order derivatives of the original field $y$ as well as the time-varying estimated field $\tilde{y}^t$. In General, it is desirable to quickly build the estimation of $E^t$, using accumulated samples over steps, so as to converge $p_i^t$ to the ideal $p_i^{t*}$ as soon as possible. Thus, larger $\frac{dE^t}{dx}$ are more preferred. Further, we argue that $y_i^t$ will gradually increase from $0$ to $y_i$ for all $i$ during SGD process. Therefore, it is implied that the proposed algorithm performs better if sensors located at steeper slopes of the original field are selected sooner, which will provide more information of the real error function, thus contribute more to subsequent learning.
\end{IEEEproof}

Other than the stochastic sampling behavior, the characteristics of the original unknown field also has a great impact on the performance of the proposed algorithm.

Here we introduce $\boldsymbol{\upsilon}=\left(\frac{|y_1|}{|y|_{\textrm{max}}},...,\frac{|y_n|}{|y|_{\textrm{max}}}\right)$, normalized by $|y|_{\textrm{max}}=\max_{i}\left({|y_1|,...,|y_n|}\right)$.

\vspace{0.1cm}
\textbf{\emph{Theorem 3:}} Compared to uniform sampling in vanilla SGD, the proposed meta-learning based algorithm improves the convergence rate if $\|\boldsymbol{\upsilon}\|_1 \ll n$.

\begin{IEEEproof}
The proof is motivated by steps in \cite{Zhao2015Stochastic}.
\begin{equation}\label{main}
\begin{split}
&\|\boldsymbol{\omega}_{t}-\boldsymbol{\omega}^{\ast}\|^{2} - \|\boldsymbol{\omega}_{t+1}-\boldsymbol{\omega}^{\ast}\|^{2}\\
=&\|\boldsymbol{\omega}_{t}-\boldsymbol{\omega}^{\ast}\|^{2} - \|\boldsymbol{\omega}_{t}-\eta_{t}(n p_{i}^{t})^{-1}\nabla L_{i}(\boldsymbol{\omega}_{t})\\
&-\eta_{t}\gamma \cdot \partial \Gamma(\boldsymbol{\omega}_{t+1})-\boldsymbol{\omega}^{\ast}\|^{2}\\
=&2\eta_{t}\langle(n p_{i}^{t})^{-1}\nabla L_{i}(\boldsymbol{\omega}_{t}), \boldsymbol{\omega}_{t}-\boldsymbol{\omega}^{\ast}\rangle\\
&+2\eta_{t}\langle\gamma \cdot \partial \Gamma(\boldsymbol{\omega}_{t+1}), \boldsymbol{\omega}_{t}-\boldsymbol{\omega}^{\ast}\rangle\\
&-\|\eta_{t}(n p_{i}^{t})^{-1}\nabla\phi_{i_t}(\boldsymbol{\omega}_{t})+\eta_{t}\gamma\cdot \partial \Gamma(\boldsymbol{w}^{t+1})\|^2.
\end{split}
\end{equation}
Invoking Lemma 2 where $L(\boldsymbol{\omega})$ is 2-strongly convex, the first term on the right-hand side in the above equation satisfies
\begin{equation}\label{first_term}
\begin{split}
&\mathbb{E}\frac{1}{n p_{i}^{t}}\big[\langle\nabla L_{i}(\boldsymbol{\omega}_{t}), \boldsymbol{\omega}_{t}-\boldsymbol{\omega}^{\ast}\rangle\\
&\quad \quad \quad-\big(L_{i}(\boldsymbol{\omega}_{t}) - L_{i}(\boldsymbol{\omega}^{\ast})+\|\boldsymbol{\omega}_{t}-\boldsymbol{\omega}^{\ast}\|^{2}\big)\big]\\
=&\langle\nabla L(\boldsymbol{\omega}_{t}), \boldsymbol{\omega}_{t}-\boldsymbol{\omega}^{\ast}\rangle
- \big(L(\boldsymbol{\omega}_{t}) - L(\boldsymbol{\omega}^{\ast})+\|\boldsymbol{\omega}_{t}-\boldsymbol{\omega}^{\ast}\|^{2}\big)\\
\geq& 0.
\end{split}
\end{equation}

Next, due to the convexity of $\Gamma(\cdot)$, we have
\begin{equation}\label{second_term}
\begin{split}
&\langle\gamma \cdot\partial\Gamma(\boldsymbol{\omega}_{t+1}),\boldsymbol{\omega}_{t}-\boldsymbol{\omega}^{\ast}\rangle \\
=&\langle\gamma \cdot\partial\Gamma(\boldsymbol{\omega}_{t+1}),\boldsymbol{\omega}_{t+1}
-\boldsymbol{\omega}^{\ast}\rangle
+\langle\gamma \cdot\partial\Gamma(\boldsymbol{\omega}_{t+1}),\boldsymbol{\omega}_{t}-\boldsymbol{\omega}_{t+1}\rangle\\
\geq& \gamma \Gamma(\boldsymbol{\omega}_{t+1})-\gamma \Gamma(\boldsymbol{\omega}^{\ast})\\
&+\langle\gamma\cdot\partial \Gamma(\boldsymbol{\omega}_{t+1}),\frac{\eta_{t}}{n p_{i}^{t}}\nabla L_{i}(\boldsymbol{\omega}_{t}) + \eta_{t}\gamma\cdot\partial \Gamma(\boldsymbol{\omega}_{t+1})\rangle.
\end{split}
\end{equation}
Substitute\eqref{first_term} and \eqref{second_term} into \eqref{main}, yielding
\begin{equation}
\begin{split}
&\mathbb{E}\left[ \|\boldsymbol{\omega}_{t} - \boldsymbol{\omega}^{\ast}\|^2 - \|\boldsymbol{\omega}_{t+1} - \boldsymbol{\omega}^{\ast}\|^2\right] \\
\geq& 2\eta_{t}\big[(L(\boldsymbol{\omega}_{t})-(\boldsymbol{\omega}^{\ast})) + \|\boldsymbol{\omega}_{t}-\boldsymbol{\omega}^{\ast}\|^2 \\
&\quad \quad + \gamma \Gamma(\boldsymbol{\omega}_{t+1}) -\gamma \Gamma(\boldsymbol{\omega}^{\ast})\big] \\
&- \eta_{t}^{2}\mathbb{E}\|(n p_{i}^t)^{-2}\nabla L_{i}(\boldsymbol{\omega}_{t})\|^2\\
&- \eta_{t}^{2}\|\gamma\cdot \partial \Gamma(\boldsymbol{\omega}_{t+1})\|^{2}.
\end{split}
\end{equation}
By taking expectation of both sides, it can be straightforwardly derived that
\begin{align}
&\mathbb{E}\left[L(\boldsymbol{\omega}_{t})-L(\boldsymbol{\omega}^{\ast}) + \lambda \Gamma(\boldsymbol{\omega}_{t+1})-\lambda \Gamma(\boldsymbol{\omega}^{\ast})\right]\notag\\
&\leq \frac{1}{2\eta_{t}}\mathbb{E}\left[\|\boldsymbol{\omega}_{t}-\boldsymbol{\omega}^{\ast}\|^2 - \|\boldsymbol{\omega}_{t+1} - \boldsymbol{\omega}^{\ast}\|^2\right] - \mathbb{E}\|\boldsymbol{\omega}_{t}-\boldsymbol{\omega}^{\ast}\|^{2}\notag\\
&~~~~+\frac{\eta_{t}}{2}\mathbb{E}\|(n p_{i}^{t})^{-1}\nabla L_{i}(\boldsymbol{\omega}_{t})\|^2.
\end{align}
Summing the above inequality over $t=1,\cdots,T$ and using $\eta_{t}=\frac{1}{2t}$, gives rise to the following:
\begin{align}
&\sum_{t=1}^{T}\mathbb{E}\left[L(\boldsymbol{\omega}_{t})+\lambda \Gamma (\boldsymbol{\omega}_{t+1})\right] - \sum_{t=1}^{T}\mathbb{E}\left[L(\boldsymbol{\omega}^{\ast})+\lambda \Gamma (\boldsymbol{\omega}^{\ast})\right] \notag\\
&\leq \sum_{t=1}^{T}t\left[\mathbb{E}\|\boldsymbol{\omega}_{t+1}-\boldsymbol{\omega}^{\ast}\|^2 - \mathbb{E}\|\boldsymbol{\omega}_{t+1} - \boldsymbol{\omega}^{\ast}\|^2\right] \notag\\
&~~~~-\sum_{t=1}^{T}\mathbb{E}\|\boldsymbol{\omega}_{t} - \boldsymbol{\omega}^{\ast}\|^2+\sum_{t=1}^{T}\frac{\eta_{t}}{2}\mathbb{E}\|(n p_{i}^{t})^{-1}\nabla L_{i}(\boldsymbol{\omega}_{t})\|^2\notag\\
&=-T \mathbb{E}\|\boldsymbol{\omega}_{t+1} - \boldsymbol{\omega}^{\ast}\|^2 + \sum_{t=1}^{T}\frac{1}{4t}\mathbb{E}\|(n p_{i}^{t})^{-1} \nabla L_{i}(\boldsymbol{\omega}_{t})\|^{2}\notag\\
&\leq \sum_{t=1}^{T}\frac{1}{4t}\mathbb{E}\|(n p_{i}^{t})^{-1}\nabla L_{i}(\boldsymbol{\omega}_{t})\|^{2} \notag\\
&=\mathbb{E}\sum_{t=1}^{T}\frac{1}{4t}\frac{1}{n^{2}}\sum_{i=1}^{n}\frac{1}{p_{i}^t}\|\nabla L_{i}(\boldsymbol{\omega}_{t})\|^2.
\label{th1}
\end{align}




Further, since
\begin{align}
\|\nabla L_{i}(\boldsymbol{\omega}_{t})\|
&=2|y_{i}-\Phi(x_{i})\boldsymbol{\omega}|\cdot \|\Phi(x_i)\|
\leq 2|y_{i}|\cdot \|\Phi_(x_i)\|\notag\\
&\leq 2|y_{i}|,
\end{align}
which shows that each $L_{i}(\boldsymbol{\omega}_{t})$ is upper-bounded by $2|y_{i}|$. Accordingly, $p_i^t \leq \frac{|y_{i}|}{\sum_{j=1}^{n}|y_{j}|}$. Together with \eqref{th1} we write
\begin{align}
\frac{1}{n^2}\sum_{i=1}^{n}\frac{1}{p_{i}^{t}}\|\nabla L_{i}(\boldsymbol{\omega}_{t})\|^2
&\leq \frac{1}{n^2}\sum_{i=1}^{n}\frac{(\sum_{j=1}^{n}|y_{j}|)4|y_{i}|^{2}}{|y_{i}|}\notag\\
&= 4\left(\frac{\sum_{i=1}^{n}|y_{i}|}{n}\right)^{2}.
\label{th2imporsgd}
\end{align}

As such, we can view vanilla SGD with uniform sampling as a special case where
 $|y_i|=|y|_{max}$,  and the distribution $p_i^t=\frac{|y|_{\textrm{max}}}{\sum_{j=1}^{n}|y|_{\textrm{max}}} = \frac{1}{n}$, thus above inequality now becomes£º
\begin{equation}
\begin{aligned}
\frac{1}{n^2}\sum_{i=1}^{n}\frac{1}{p_{i}^{t}}\|\nabla L_{i}(\boldsymbol{\omega}_{t})\|^2
&\leq \frac{1}{n^2}n\cdot n\cdot 4|y|^2_{max} \\
&=4 |y|^2_{max}
\end{aligned}
\label{th2sgd}
\end{equation}
Taking ratio between \eqref{th2sgd} and \eqref{th2imporsgd}, yields
\begin{equation}
\frac{|y|_{\textrm{max}}^2}{\left(\frac{(\sum_{i=1}^{n}|y_i|)^{2}}{4n^{2}}\right)}=\frac{n^2|y|_{\textrm{max}}^2}{\left(\sum_{i=1}^{n}|y_i|\right)^2},
\end{equation}
which implies the improvement on convergence rate, especially when $\|\boldsymbol{\upsilon}\|_1 = \sum_{i=1}^{n}\frac{|y_i|}{|y|_{\textrm{max}}}
 \ll n$.
\end{IEEEproof}

\emph{Remark:} Theorem 3 indicates that the performance gain provided by the proposed algorithm is more obvious when the field contains less but distinguishing features. For better illustration, we plot 3 truncated windows of the same length under one-dimensional case. The X axis represents the one-dimensional location, the Y axis stands for the corresponding field value. Note that the peaks within are all set to the same height, thus field (a) and (b) only differ in the number of features, while the (b) and (c) differ in the shape of the feature, i.e., the spread. Comparing $\boldsymbol{\upsilon}$s
of each field, obviously, $\|\boldsymbol{\upsilon}_b\|_1<\|\boldsymbol{\upsilon}_a\|_1<\|\boldsymbol{\upsilon}_c\|_1$, hence proposed algorithm tends to advances most in field (b).
 \begin{figure}[!ht]
 \centering
 \includegraphics[width=0.48\textwidth]{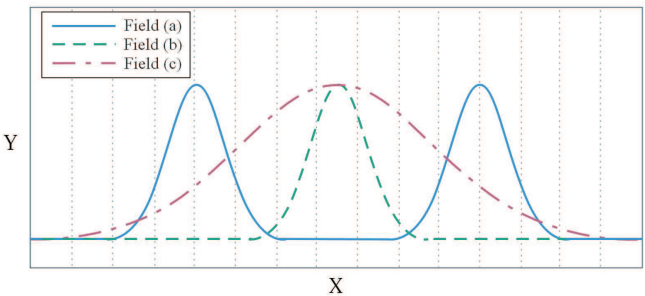}
 \caption{3 truncated windows of the same length in one-dimensional fields.}
 \label{threecases}
 \end{figure}

\section{Experiments}

In this section, we evaluate the performance of the proposed algorithm and compare it with the conventional ones in terms of convergence properties and communication cost.

\subsection{Experiment Setup}

The simulation is conducted for a WSN with $n$ distributedly deployed sensors measuring some unknown environmental quantity. Practically, it can stand for typical environment monitoring scenarios in WSN-based IoT. For instance, the indoor/outdoor temperature field of a residence needs to be estimated through deployed sensors, in order to enable a variety of ``Smart Home'' applications, such as automated air-conditioning, floor heating, etc.

\subsubsection{Task Generation}
 Here for useful insights and simplicity, we only take the one-dimensional scenario for illustration. Note two or more dimensional case can be directly extended.

The 1-D spatial domain is confined to $x \in [-5,5]$, and the target field function $f(x)$ is supposed to be a weighted sum of $K=50$ potential Gaussian kernels with equally-spaced centers $\{c_k\}$ and identical width $\beta=0.4$. To introduce certain degree of sparsity as well as randomness of the unknown field, $\kappa$ out of $K$ entries of the parameter vector $\boldsymbol{\omega}\in \mathbb{R}^{K \times 1}$ are randomly chosen to be nonzero, i.i.d Gaussian variables. Intuitively, larger $\kappa$ will lead to more complex and fluctuated field.

Second, to simulate the sensing scenario, we assume a total of $n=500$ sensors are randomly deployed throughout the field, and each sensor gets its observation according to $y_i=f(x_i)+n_i$, where $n_i\sim N(0,\sigma_n^2)$ is the zero-mean Gaussian noise with variance $\sigma_n=0.1$.
Now, a FC begins to access these sensors one at a time. At each access, the FC collects an observation and stores it in the buffer. Meanwhile the sampled observations $\{(x_1,y_1),...,(x_n,y_n)\}$ are utilized to train the field model.

As explained in Section \uppercase\expandafter{\romannumeral2}, this particular field reconstruction problem corresponds to finding the optimal parameter satisfying:
\begin{equation}
\boldsymbol{\omega}^{*} = \arg \min_{\boldsymbol{\omega}} \frac{1}{n} \sum_{i=1}^{n} \left(\Phi(x_i)\boldsymbol{\omega}-y_i\right)^2 + \gamma \|\boldsymbol{\omega}\|_1
\label{goal2}
\end{equation}
based on all potential measurements $\{(x_1,y_1),...,(x_n,y_n)\}$.

The above field are repeatedly generated for 200 times and the above procedure runs for the same number of times as well, which then be used to meta-train an optimal sampling policy in the sequel.

\subsubsection{Strategies}
\begin{figure}[!ht]
\centering
\includegraphics[width=0.45\textwidth]{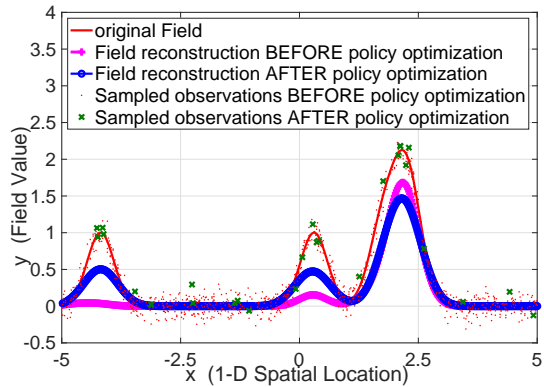}
\caption{Field reconstruction and selected observations before/after policy optimization.}
\label{before_after_poli_opt}
\end{figure}

 Due to the real-time processing nature of the above task, the Proximal SGD mechanism in \eqref{recursive} is leveraged to solve the problem in \eqref{goal2}, where the gradient of each step is reflected by each individual observation and the learning rate $\eta_t$ and the penalty factor $\gamma$ are set to $1/(2t)$ and 0.08, respectively. On this basis, the only factor that influences the task performance lies in the sampling strategy along the process. Here we compare the proposed meta-learning based sampling with its conventional counterpart, the uniform sampling, by appling them to the above SGD framework. In detail, the two are described below:
\begin{itemize}
\vspace{0.1cm}
	\item \textbf{Meta-based Sampling}: The optimal sampling policy based on the proposed two-layer learning and sensing algorithm, meta-trained upon the 200 tasks generated above. The meta-training process on a particular task is listed in Algorithm 2. To avoid over-fitting, we freeze the parameter $\Theta$ after $L=10$ episodes of training on a single task, and then evaluate its performance on the other tasks. We choose the policy that achieves the best expected reward to be the final optimal policy. Fig. \ref{before_after_poli_opt} qualitatively compares the field reconstruction performance before and after policy optimization. The X-axis represents the one-dimensional spatial location, $x\in [-5,5]$, and Y-axis denotes the corresponding field value at each location. As shown, with the red solid line being the originally generated field, the blue curve, representing the field reconstruction AFTER policy optimization, is apparently much more accurate than that BEFORE, yet it enjoys a much lower communication cost (or sample numbers). More interestingly, most of the sample sensed (the green crosses in the figure) resides nearby the abrupt changes of the curve such as peaks or valleys which capture the most critical features of the field, whereas the FC tends to allocate less sensing effort to those in the smooth region.
\vspace{0.1cm}
    \item \textbf{Uniform Sampling}: The Proximal SGD algorithm with uniform sampling, i.e.,
    $i_{t+1}$ is randomly picked from $1,2,...,n$.
\end{itemize}
\begin{algorithm}[H]
\label{meta-training}
\caption{Meta-training flow on a particular task}
\begin{algorithmic}
\REQUIRE Training data $D={(x_i,y_i)}_{i=1,...,n}$, maximum iteration number $T$, episode number $L$, and discount factor $\lambda$.
\STATE Initialize the sampling policy $\pi_{\Theta}^{(0)}$ or equivalently, $P_{\Theta}^{(0)}(a|s)$ arbitrarily
\FOR {each episode $l=0,1,2,...,L$}
     \STATE Initialize base training model, i.e., $\boldsymbol{\omega}_0$
     \FOR {t=1,2,...,T}
     	\STATE Sample action $a_t$ according to current policy $\pi_{\Theta}^{(l)}$, and update the base training model based on the selected data $D_{a_t}$, meanwhile receive reward $r_t$.
     \ENDFOR
     \STATE For each trajectory $\{s_{0},a_{1},r_{1},\cdots,s_{T-1},a_{T},r_{T}\}$
     \FOR {$t=1$ to $T-1$}
		\STATE $\theta\leftarrow \theta + \alpha \nabla_{\theta} \log \pi_{\theta} (s_{t},a_{t})v_{t}$
	 \ENDFOR
\ENDFOR \\
\ENSURE $\pi_{\Theta}^{(L)}$
\end{algorithmic}
\end{algorithm}
%
%
%
%

\subsection{Experiment Results}
\begin{figure}[!ht]
\centering
\subfigure[$\kappa=4$]{
\includegraphics[width=0.48\textwidth]{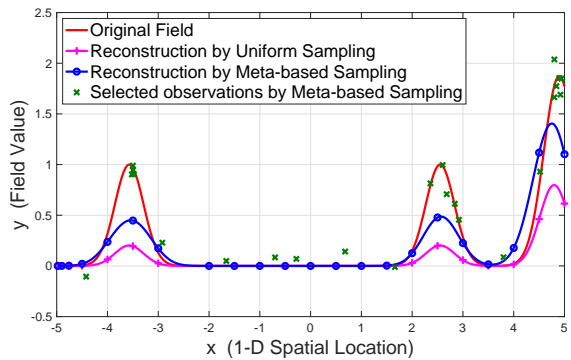}
\label{ka4reconstruction}}
\subfigure[$\kappa=8$]{
\includegraphics[width=0.48\textwidth]{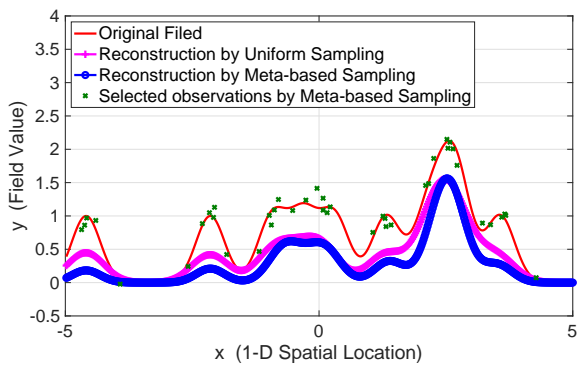}
\label{ka8reconstruction}}
\subfigure[$\kappa=20$]{
\includegraphics[width=0.48\textwidth]{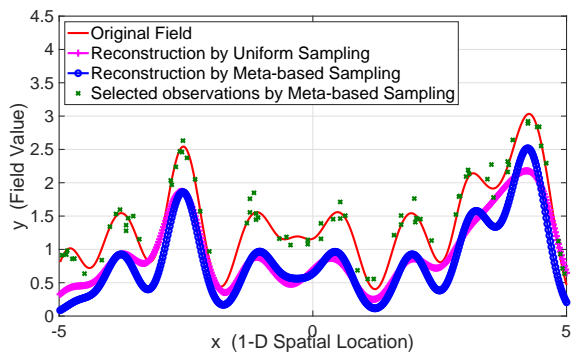}
\label{ka20reconstruction}}
\caption{Comparison of samples' locations and reconstruction performance.}
\end{figure}

 We directly apply the meta-trained policy, as well as the uniform sampling strategy to 500 testing tasks with various $\kappa$, i.e., to the case where the field has changed, either to become more fluctuated or smoother. From Fig. \ref{ka4reconstruction} to Fig .\ref{ka20reconstruction}, it is observed that the meta-learner is capable of providing the base-learner with more crucial and informative training samples, instead of those redundant ones, thus yielding a much better reconstruction performance.

\begin{figure}[!ht]
\centering
\includegraphics[width=0.45\textwidth]{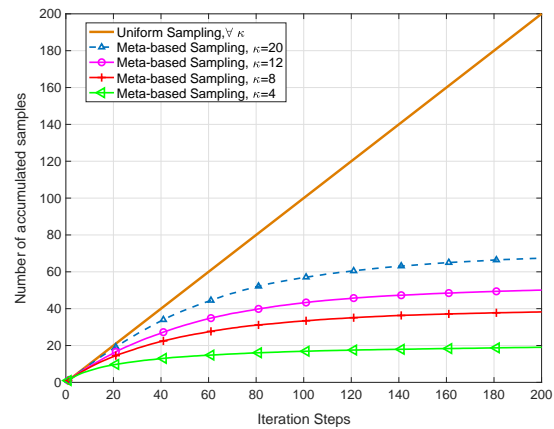}
\caption{Comparison of accumulated samples at FC along the training process, under different $\kappa$.}
\label{ka481220_numberobserv}
\end{figure}

 In terms of communication cost, as shown in Fig. \ref{ka481220_numberobserv}, as time goes, the number of samples sensed in the proposed meta-learning based sampling policy grows much more slowly in comparison with that of the uniform sampling (note that the benchmark importance sampling always needs to evaluate all samples over the field). Moreover, it gradually converges to some upper bounds that increase with $\kappa$, indicating that no more samples are needed to meet certain fusion accuracy. It can be interpreted that the meta-learning based sampling policy in fact enables the FC to adaptively shift between exploration and exploitation. The former tends to explore the unobserved portion of the field, whereas the latter makes use of the existing samples without inducing extra communication cost. This way, it ``intelligently'' decides whether the number of samples is enough or not. Intuitively, the number of samples may increase accordingly given a more complex field with larger $\kappa$, to guarantee the required reconstruction performance.

 We further compare the convergence rate between different sampling strategies, by evaluating the averaged mean squared error (MSE), i.e., the first loss component in \eqref{goal2} at each iteration step. Note that here an additional strategy, known as the ``Importance-based Sampling'' \cite{Zhao2015Stochastic}, is also involved for reference purpose. Specifically, it samples ideally according to \emph{importance} over the entire dataset (whether observed or not), as reflected by its norm of gradient, namely,
\begin{equation}
 i_{t+1}\sim p_{i_{t+1}}=\frac{\|\nabla L_{i_{t+1}}(\boldsymbol{\omega}_t)\|}{\sum_{j=1}^{n}  \|\nabla L_j(\boldsymbol{\omega}_t)\|}
\end{equation}

\begin{figure}[!ht]
\centering
\subfigure[$\kappa=4$]{
\label{ka4converg} 
\includegraphics[width=0.4\textwidth]{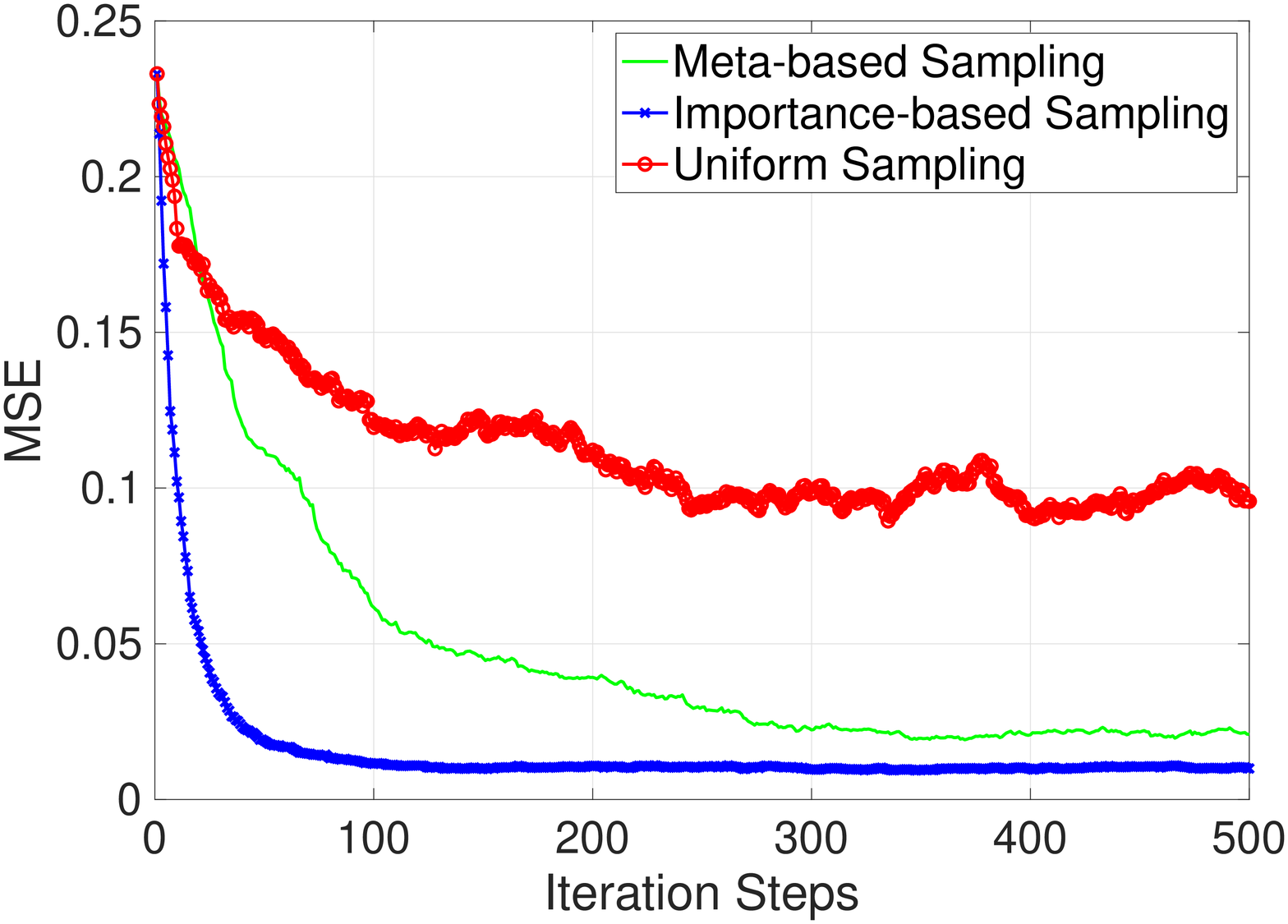}}
\subfigure[$\kappa=8$]{
\label{ka8converg} 
\includegraphics[width=0.4\textwidth]{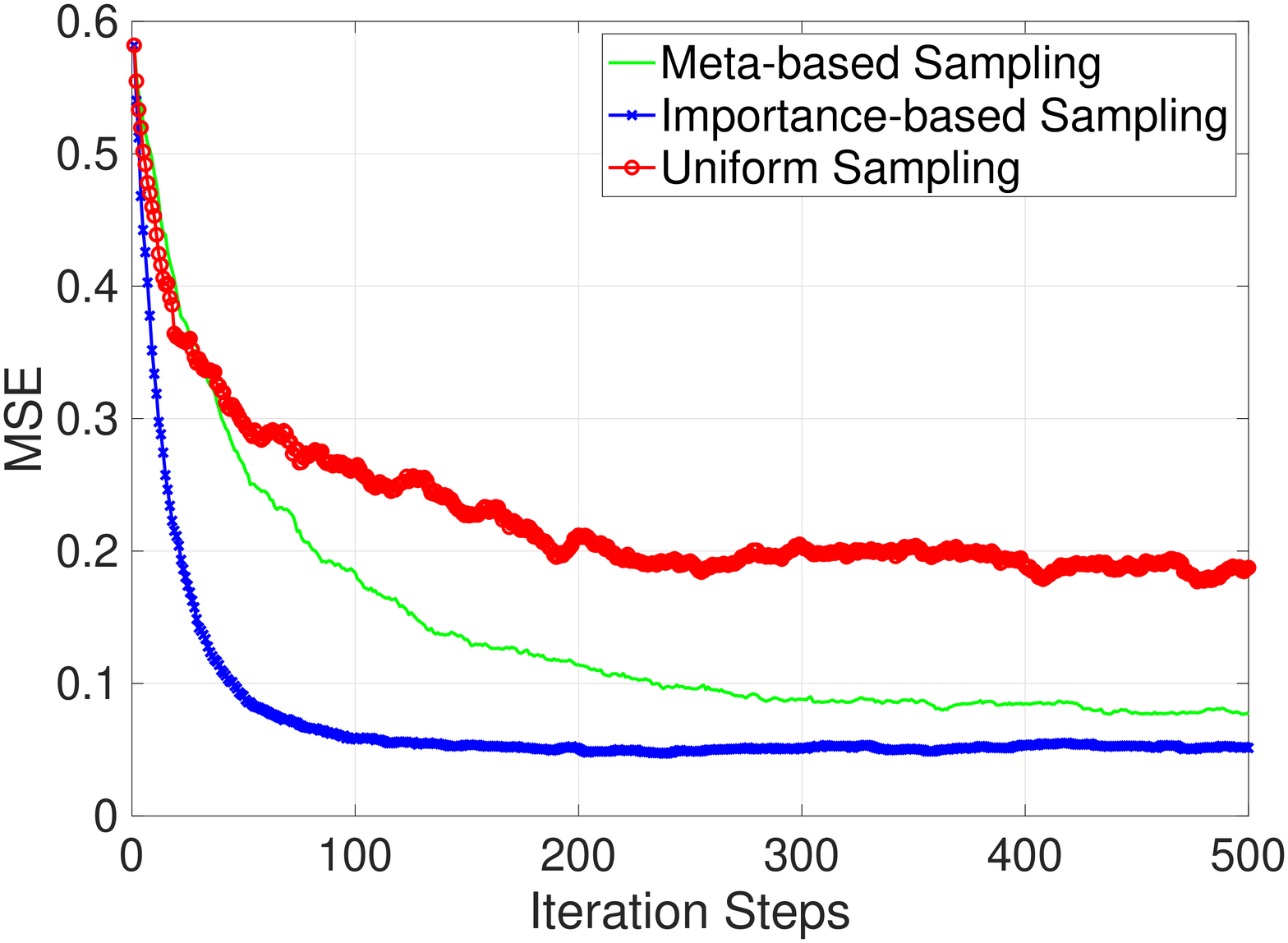}}
\subfigure[$\kappa=12$]{
\label{ka12converg} 
\includegraphics[width=0.4\textwidth]{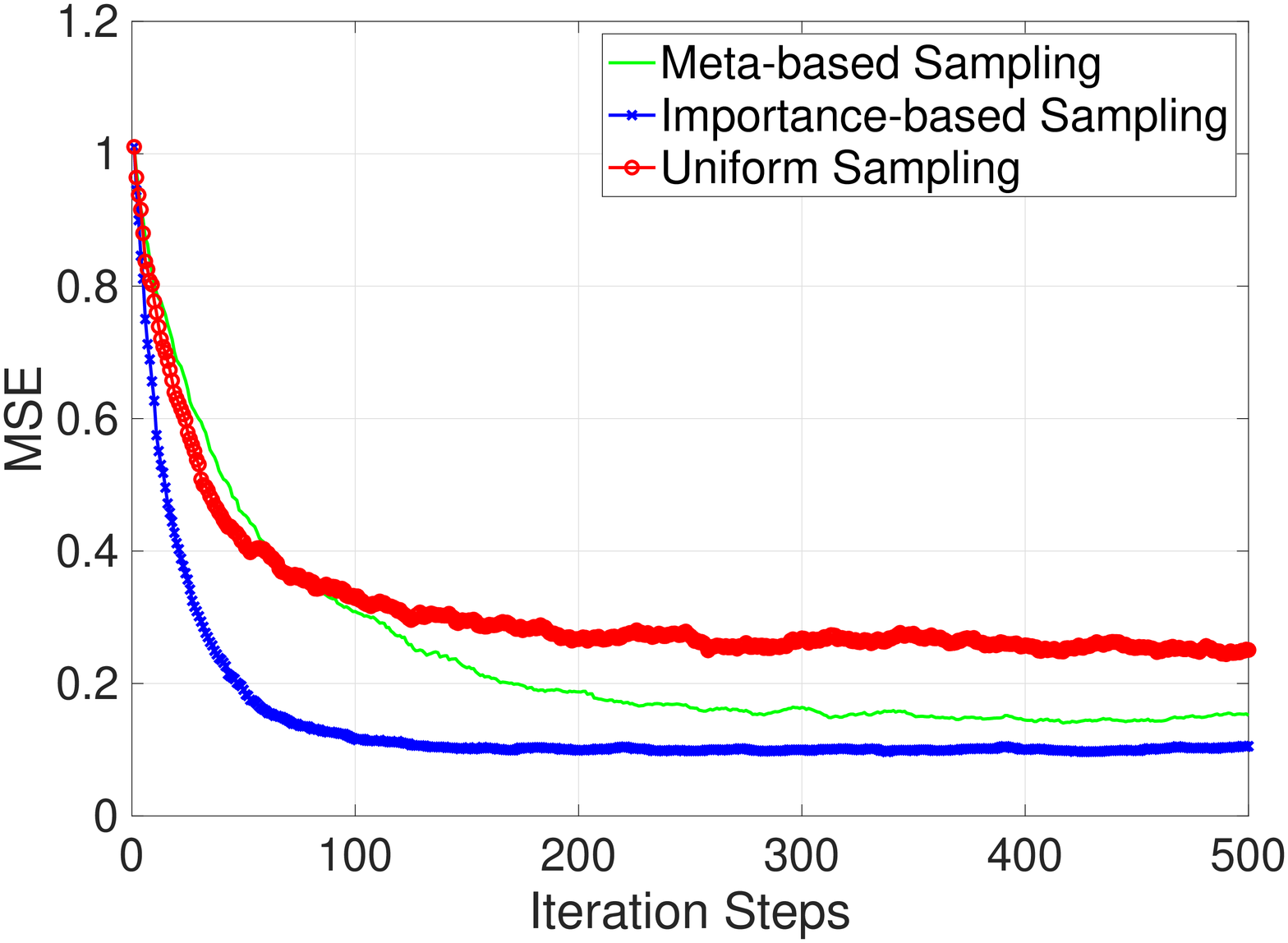}}
\subfigure[$\kappa=20$]{
\label{ka20converg} 
\includegraphics[width=0.4\textwidth]{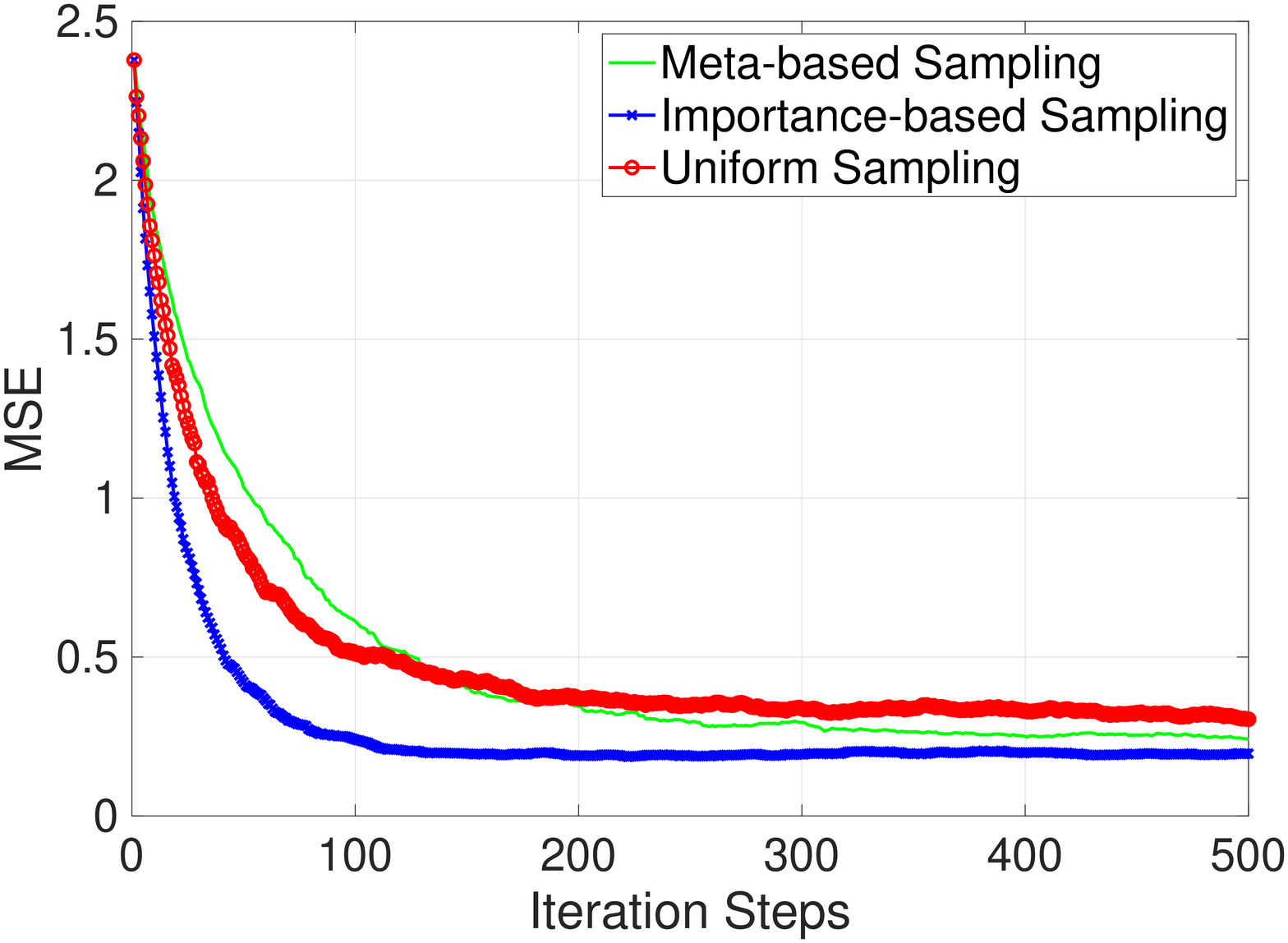}}
\caption{Convergence comparison between meta-based and uniform sampling under different $\kappa$.}
\label{ka_converg} 
\end{figure}

 As shown in Fig. \ref{ka_converg}. On the one hand, the performance of the meta-based sampling scheme is upper-bounded by that of the importance-based sampling, where the gap in-between stands for the incremental learning process of gradually accumulating information of the field. In this sense, we conclude that the meta-based sampling can reach an sub-optimal convergence, but without suffering the communicational expense of collecting all the samples beforehand.

\begin{figure}[!ht]
\centering
\includegraphics[width=0.45\textwidth]{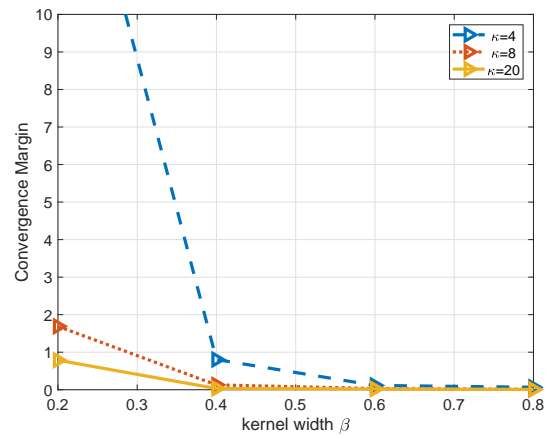}
\caption{Convergence victory margin over uniform sampling under different $\kappa$ and $\beta$.}
\label{convergence_marginkawidth}
\end{figure}

On the other hand, it beats the uniform sampling used in vanilla SGD with a faster average loss dropping rate and a lower prediction error and variance,
Nonetheless, as shown in Fig. \ref{convergence_marginkawidth},  this kind of convergence victory margin perishes with increasing $\kappa$, i.e., the actual number of effective kernels and/or $\beta$, the kernel width. The is because that larger $\kappa$ usually induces more drastic fluctuations to the field, while larger $\beta$ means a smoother field. In both cases, the importances of all potential locations tend to be equal, which makes the meta-based sampling scheme gradually boils down to the uniform sampling. This also verifies the remark on Theorem 3.

\section{Conclusion and Future Works}
In the paper, we study the WSN-based field sensing and reconstruction problem.
we establish a two-layer learning framework based on reinforcement learning, and present the detailed design for an adaptive sampling policy which
can actively determine the most informative sensing location and thus significantly reduce the communication cost. Numerical results show that the
algorithm brings a remarkable improvement in reconstruction performance and efficiency compared to conventional ones, and it also exhibits
good robustness to both information dynamics and the variation of field parameters. However, there are still many interesting problems left open on this topic.

For example, we aim to further enhance the online learning framework to make it more adaptive to the dynamic changes of field features, and to quantify the tradeoff the computational complexity of learning with the sensing and / or communication cost in this framework. Or more fundamentally, we want to derive the closed-form results of how many samples are needed to reconstruct the field by using this framework. We believe these problems are of particular importance and will leave them as our future work.


\bibliographystyle{IEEEtran}

\begin{thebibliography}{1}
\bibitem{Joshi2009Sensor} S.~Joshi and S.~Boyd, ``Sensor selection via convex optimization,'' \emph{IEEE Trans. on Signal Processing}, vol.~57, no.~2, pp. 451--462, 2009.

\bibitem{Ghassemi2011Separable} F.~Ghassemi and V.~Krishnamurthy,
    ``Separable approximation for solving the
  sensor subset selection problem,'' \emph{IEEE Trans. on Aerospace and
  Electronic Systems}, vol.~47, no.~1, pp. 557--568, 2011.

\bibitem{Sebastiani2000Maximum} P.~Sebastiani and P.~Henry, ``Maximum
    entropy sampling and optimal bayesian
  experimental design,'' \emph{J. of the Royal Statistical Society}, vol.~62,
  no.~1, pp. 145--157, 2000.

\bibitem{Paninski2005Asymptotic} L.~Paninski, \emph{Asymptotic Theory of
    Information-Theoretic Experimental
  Design}.\hskip 1em plus 0.5em minus 0.4em\relax MIT Press, 2005.

\bibitem{Willett2004Backcasting} R. Willett, A. Martin and R. Nowak,
    ``Backcasting: Adaptive sampling for sensor
  networks,'' in \emph{Proc. of Int. Symposium on Information Processing in Sensor
  Networks}, 2004, pp. 124--133.

\bibitem{Nowak2003Boundary} R.~Nowak and U.~Mitra, ``Boundary estimation in
    sensor networks: Theory and
  methods,'' \emph{IPSN}, vol. 2634, pp. 80--95, 2003.

\bibitem{Rui2005Faster} C. Rui, R. Willett and R. Nowak, ``Faster rates in
    regression via active learning,''
  in \emph{Proc. of Int. Conf. on Neural Information Processing Systems}, 2005, pp.
  179--186.

\bibitem{Rui2008Active} C.~Rui and R.~Nowak, \emph{Active Learning and
    Sampling}.\hskip 1em plus 0.5em
  minus 0.4em\relax Springer US, 2008.

\bibitem{Suh2016Mobile} J.~Suh, S.~You, S.~Choi, and S.~Oh, ``Vision-Based
    Coordinated Localization for Mobile Sensor Networks,'' \emph{IEEE Trans.
    on Automation Science and Engineering}, vol.~13, no.~2, pp. 611--620,
    2016.

\bibitem{Nguyen2015Information} L. Nguyen, S. Kodagoda, R. Ranasinghe and G.
    Dissanayake, ``Information-driven adaptive
  sampling strategy for mobile robotic wireless sensor network,'' \emph{IEEE
  Trans. on Control Systems Technology}, vol.~24, no.~1, pp. 372--379, 2015.

\bibitem{Grasso2016Dynamic} R. Grasso, P. Braca and M. S. Greco, ``Dynamic
    underwater glider network for
  environmental field estimation,'' \emph{IEEE Trans. on Aerospace and
  Electronic Systems}, vol.~52, no.~1, pp. 379--395, 2016.

\bibitem{Popa2006Adaptive} D. O. Popa, M. F. Mysorewala and F. L. Lewis,
    ``Adaptive
    sampling using non-linear ekf
  with mobile robotic wireless sensor nodes,'' in \emph{Proc. of Int. Conf. on Control,
  Automation, Robotics and Vision}, 2006, pp. 1--6.

\bibitem{Popa2007EKF}
------, ``Ekf-based adaptive sampling with mobile robotic sensor nodes,'' in
  \emph{Proc. of IEEE Int. Conf. on Intelligent Robots and Systems}, 2007, pp.
  2451--2456.

\bibitem{Mysorewala2012A} M. Muhammad, C. Lahouari and D. O. Popa, ``A
    distributed
    multi-robot adaptive sampling
  scheme for the estimation of the spatial distribution in widespread fields,''
  \emph{Eurasip Journal on Wireless Communications and Networking}, vol. 2012,
  no.~1, pp. 1--19, 2012.

\bibitem{Buhmann2003RBF} Buhmann, Martin Dietrich (2003). Radial basis
    functions: theory and implementations. Cambridge University Press. ISBN
    978-0511040207.

\bibitem{Robbins1951A} H.~Robbins and S.~Monro, ``A stochastic approximation
    method,'' \emph{Annals of
  Mathematical Statistics}, vol.~22, no.~3, pp. 400--407, 1951.

\bibitem{Zhao2015Stochastic} P.~Zhao and T.~Zhang, ``Stochastic optimization
    with importance sampling,''
  \emph{arXiv:1401.2753v2 [stat.ML]}, pp. 1--9, 2015.

\bibitem{Needell2016Stochastic} D. Needell, N. Srebro and R. Ward,
    ``Stochastic
    gradient descent, weighted sampling,
  and the randomized kaczmarz algorithm,'' \emph{Mathematical Programming},
  vol. 155, no. 1-2, pp. 549--573, 2016.

\bibitem{Andrychowicz2016Learning} M. Andrychowicz, M. Denil and S. Gomez,
    ``Learning to learn by gradient descent by
  gradient descent,'' \emph{arXiv:1606.04474v2 [cs.NE]}, 2016.

\bibitem{Ravi2017Optimization} S.~Ravi and H.~Larochelle, ``Optimization as
    a model for few-shot learning,''
  in \emph{Proc. of International Conference on Learning Representations (ICLR)}, 2017.

\bibitem{Li2016Learning} K.~Li and J.~Malik, ``Learning to optimize,''
    \emph{arXiv:1606.01885v1 [cs.LG]}, 2016.

\bibitem{Li2017Learning}
------, ``Learning to optimize neural nets,'' \emph{arXiv:1703.00441v2 [cs.LG]}, 2017.

\bibitem{Williams1992Simple} R.~Williams, ``Simple statistical
    gradient-following algorithms for
  connectionist reinforcement learning,'' \emph{Machine Learning}, vol.~8, no.
  3-4, pp. 229--256, 1992.

\bibitem{Papa2015Adaptive} G. Papa, P. Bianchi and S. Clemencon, ``Adaptive
    sampling for incremental
  optimization using stochastic gradient descent,'' in \emph{Proc. of Int. Conf. on
  Algorithmic Learning Theory}, 2015, pp. 317--331.

\bibitem{Zhang2004Solving} T.~Zhang, ``Solving large scale linear prediction
    problems using stochastic
  gradient descent algorithms,'' in \emph{Proc. of Int. Conf. on Machine Learning.
  Omnipress}, 2004, p. 116.
\end{thebibliography}

\end{document}